\documentclass[preprint]{aastex}

\slugcomment{Accepted by the ApJL}

\shorttitle{HOT CHANNEL BEING MAGNETIC FLUX ROPE}

\shortauthors{Song et al.}

\begin{document}
\title{EVIDENCE OF THE SOLAR EUV HOT CHANNEL AS A MAGNETIC FLUX ROPE FROM REMOTE-SENSING AND IN-SITU OBSERVATIONS}
\author{H.Q. SONG\altaffilmark{1}, Y. CHEN\altaffilmark{1}, J. ZHANG\altaffilmark{2}, X. CHENG\altaffilmark{3}, B. Wang\altaffilmark{1}, Q. HU\altaffilmark{4}, G. LI\altaffilmark{4}, AND Y.M. WANG\altaffilmark{5}}

\affil{1 Shandong Provincial Key Laboratory of Optical Astronomy
and Solar-Terrestrial Environment, and Institute of Space
Sciences, Shandong University, Weihai, Shandong 264209, China}
\email{hqsong@sdu.edu.cn}

\affil{2 School of Physics, Astronomy and Computational Sciences,
George Mason University, Fairfax, VA 22030, USA}

\affil{3 School of Astronomy and Space Science, Nanjing
University, Nanjing, Jiangsu 210093, China}

\affil{4 Department of Space Science and CSPAR, University of
Alabama in Huntsville, Huntsville, AL 35899, USA}

\affil{5 Key Laboratory of Geospace Environment, University of
Science and Technology of China, Chinese Academy of Sciences
(CAS), Hefei, Anhui 230026, China}

\begin{abstract}
Hot channels (HCs), high temperature erupting structures in the
lower corona of the Sun, have been proposed as a proxy of magnetic
flux ropes (MFRs) since their initial discovery. However, it is
difficult to make definitive proof given the fact that there is no
direct measurement of magnetic field in the corona. An alternative
way is to use the magnetic field measurement in the solar wind
from in-situ instruments. On 2012 July 12, an HC was observed
prior to and during a coronal mass ejection (CME) by the AIA
high-temperature images. The HC is invisible in the EUVI
low-temperature images, which only show the cooler leading front
(LF). However, both the LF and an ejecta can be observed in the
coronagraphic images. These are consistent with the high
temperature and high density of the HC and support that the ejecta
is the erupted HC. In the meanwhile, the associated CME shock was
identified ahead of the ejecta and the sheath through the COR2
images, and the corresponding ICME was detected by \textit{ACE},
showing the shock, sheath and magnetic cloud (MC) sequentially,
which agrees with the coronagraphic observations. Further, the MC
contained a low-ionization-state center and a
high-ionization-state shell, consistent with the pre-existing HC
observation and its growth through magnetic reconnection. All of
these observations support that the MC detected near the Earth is
the counterpart of the erupted HC in the corona for this event.
Therefore, our study provides strong observational evidence of the
HC as an MFR.

\end{abstract}


\section{INTRODUCTION}
Coronal mass ejections (CMEs), the most energetic eruption in the
solar system, can cause geomagnetic activities when they interact
with the Earth's magnetosphere (Gosling et al. 1991; Zhang et al.
2003; Zhang et al. 2007), which will affect and even damage the
satellites, power grids and GPS navigation systems. Nowadays, the
solar physics community has reached a general consensus that CMEs
are initiated by the eruption of magnetic flux ropes (MFRs) (e.g.,
Chen 2011; Cheng et al. 2013, 2014a). The in-situ detections of
magnetic clouds (MCs) contained in the interplanetary coronal mass
ejections (ICMEs) (Burlaga et al. 1981) provide a direct evidence
of the MFR existence. In the outer corona, at least $\sim$40\% of
coronagraphic observations of CMEs show an apparent MFR geometry
(Vourlidas et al. 2013), further supporting that the CME
structures contain the MFRs. It is believed that MFRs should exist
in the inner corona, either they are formed before (Patsourakos et
al. 2013) or during (Song et al. 2014a) the eruptions. However, it
is still an open question that what structures really depict the
MFRs in the inner corona as we do not have the direct measurements
of the coronal magnetic fields yet.

In addition to sigmoids (Titov \& D\'emoulin 1999; McKenzie \&
Canfield 2008) and coronal cavities (Wang \& Stenborg 2010), which
are widely regarded as proxies of MFRs in the inner corona, Zhang
et al. (2012) first reported and suggested that a new
observational line, hot channels (HCs), is the MFR that exists in
the inner corona. HCs refer to the high temperature structure
revealed first by the Atmospheric Imaging Assembly (AIA) 131 \AA \
passband (sensitive to $\sim$10 MK) and are invisible in the
cooler temperature images (e.g., AIA 171 \AA \ passband, sensitive
to $\sim$0.6 MK ). An HC will appear as a hot blob when observed
along its axis due to the projection effect (Cheng et al. 2011;
Patsourakos et al. 2013; Song et al. 2014a, 2014b). Since the
discovery of HCs, some evidence has been reported to support that
they might be the MFRs. For instance, Cheng et al (2014a)
presented an HC with helical threads winding around an axis, which
indicates the intrinsic helical structure of HC; Cheng et al.
(2011) found that an HC can grow during the eruption, consistent
with the MFR growth process in the classic magnetic reconnection
scenario. This further supports that an HC is the MFR structure;
Cheng et al. (2014b) reported that an HC was cospatial with a
prominence, while the HC top separated from that of the prominence
during the eruption, which offered a new evidence that the HC is
an MFR as it is generally accepted that a prominence can exist at
the dip of an MFR (Rust \& Kumar 1994).

Though many studies indicate that HCs are the MFRs, the direct
observational evidence remains lacking. As MFRs are the volumetric
plasma structure with magnetic field lines wrapping around an
axis, the substantial evidence that supports a structure being the
MFR should be based on its measurements of the magnetic fields.
However, no reliable measurements of the coronal magnetic fields
are available at present as mentioned. Therefore, we anticipate to
study the magnetic fields of HCs through the in-situ detections of
their interplanetary counterparts, which might provide strong
observational evidence that HCs have the structure of helical
fields.

In this letter, a CME induced by an HC eruption and its associated
ICME are investigated. The ICME contains a typical MC structure,
which should be the interplanetary counterpart of the erupted HC.
Therefore, we provide strong observational evidence that HCs
correspond to the MFRs. The instruments and methods are introduced
in Section 2, and the relevant observations and results are
described in Section 3. Section 4 presents the discussion, and
Section 5 is a summary.

\section{INSTRUMENTS AND METHODS}

The eruption process in the lower corona was recorded by the AIA
telescope (Lemen et al. 2012) on board the \textit{Solar Dynamic
Observatory (SDO)} and the Extreme Ultraviolet Imager (EUVI) on
board the \textit{Solar Terrestrial Relations Observatory
(STEREO)} A and B from three different perspectives. AIA has four
telescopes to observe the solar atmosphere through 10 narrow UV
and EUV passbands with a high cadence (12 s), a high spatial
resolution (1.2$''$), and a large FOV (1.3 $R_\odot$). The EUVI
provides the solar EUV images at four wavelengths. The related CME
was observed with the white light coronagraphs on board
\textit{STEREO}, including COR1 (FOV: 1.4-4 R$_\odot$) and COR2
(FOV: 2.5-15 R$_\odot$) (Howard et al. 2008). Near the earth, the
ICME was detected by the \textit{Advanced Composition Explorer
(ACE)} satellite. We used the in-situ data from MAG (Smith et al.
1998), SWEPAM (McComas et al. 1998) and SWICS (Gloeckler et al.
1998) to analyze the solar wind magnetic field and plasma
properties. The soft X-ray (SXR) data are from the
\textit{Geostationary Operational Environment Satellite (GOES)}.
\textit{GOES} provides the integrated full-disk SXR emission from
the Sun, which are used to define the magnitude, onset time, and
peak time of solar flares.

The AIA has a broad temperature coverage from 0.6 to 20 MK
(O'Dwyer et al. 2010; Del Zanna et al. 2011; Lemen et al. 2012),
and is ideal for constructing the differential-emission-measure
(DEM) models of the coronal structures (e.g., Cheng et al. 2012).
A DEM-weighted average temperature is used to analyze the HC's
initial thermal evolution (see Cheng et al. 2012; Song et al.
2014b).

\section{OBSERVATIONS AND RESULTS}

\subsection{Overview of the Eruption}
On 2012 July 12, \textit{GOES} recorded an X1.4 class SXR flare,
which located at the heliographic coordinates S17W08 (NOAA 11520)
from the Earth perspective. The corresponding SXR flux rose
gradually from $\sim$14:50 UT, then started to increase rapidly at
$\sim$16:10 UT and peaked at $\sim$16:49 UT.

Cheng et al. (2014c) have analyzed this event and concluded that
there were a high-lying MFR (a diffuse and elongated HC) and a
low-lying MFR (a sigmoid) coexisting above the same polarity
inversion line (PIL) of the active region for 2 hr prior to the
eruption, which formed a double-decker MFR system. Just the
high-lying MFR erupted and the associated MC arrived at
\textit{ACE} at July 15 06:00 UT. This CME propagation process
from the Sun to near the Earth has been reported by M\"ostl et al.
(2014) and Hess \& Zhang (2014). Shen et al. (2014) presented a
data-constrained 3-D (three-dimensional) magnetohydrodynamic
simulation for the CME propagation in the corona and
interplanetary space, consistent well with the observations.
Therefore, there is no doubt that the ICME detected with the
in-situ data is corresponding to the CME induced by the HC
eruption. We revisit this event to investigate whether the
detected MC near the Earth is the counterpart of the erupted HC in
the corona.

\subsection{The HC Eruption and Its Associated CME}

When the high-lying HC eruption took place, \textit{STEREO} A and
B were 120$^{\circ}$ west and 115$^{\circ}$ east of the Earth with
distances of 0.96 and 1.02 AU, respectively. As the separation
angle is close to 90$^{\circ}$, \textit{STEREO} A (B) provides the
southeast (southwest) limb view of the eruption as shown in right
(left) panels of Figure 2, while \textit{SDO} presents the disk
observation of the active region as presented in Figure 1.

Figure 1(a) is the profile of \textit{GOES} SXR 1-8 \AA \ flux.
The two vertical black lines show the corresponding observation
time of the middle and bottom panels, where the AIA 94 \AA\ , 131
\AA \ and the temperature map obtained through the DEM method are
presented in the left, middle and right panels, respectively. As
mentioned, a double-decker MFR system existed in this event. The
blue (purple) dotted lines in the middle and bottom panels depict
the low (high) lying MFR. Note in panels (b3) and (c3), the purple
is replaced with the white. The high lying HC is diffuse and less
bright than the flare region, so only the animation of the AIA 94
\AA \ and 131 \AA \ images accompanying Figure 1 permits the
appreciation of the HC's shape and dynamics (also see Cheng et al.
2014c). Usually, the background emissions are less outside the
solar disk and HCs are clearer in the limb events (Zhang et al.
2012; Song et al. 2015). The temperatures of the HC and sigmoid
are around 5 MK before the SXR flux began to increase slowly
(14:50 UT). After the flare onset, both structures are heated, and
show obvious temperature enhancement, especially the sigmoidal
region (16:10 UT). As the HC in this event became more diffuse and
moved out of the AIA FOV soon after the flare impulsive phase
onset, further studies about its growth and heating processes are
not available. However, it is reasonable to anticipate that the HC
will grow up as more poloidal magnetic fluxes injected during the
magnetic reconnection (Cheng et al. 2011). In the meantime, its
temperature should increase during the flare impulsive phase as
expected like a failed HC eruption event (Song et al. 2014b).

The erupting HC shows a writhed morphology with its dominant part
lying horizontally from the Earth perspective. Therefore, the HC
was observed edge on by \textit{STEREO} as shown in Figure 2.
However, just the compressed cooler leading front (LF) was
recorded (see panels (a) and (b)) by the EUVI 195 \AA \ images
(sensitive to $\sim$1.5 MK). As the EUVI does not have the
passbands sensitive to high temperatures, the ejecta is invisible
through its images (e.g., 195 \AA, see the animation (a)
accompanying Figure 2), which is consistent with the HC's high
temperature. The corresponding CME can be well observed by COR1 as
displayed in panels (c) and (d). The coronagraphic images can show
the LF and the ejecta/HC clearly as depicted with the arrows,
because both the compressed region and the HC have the higher
density in the corona compared to the background regions (Cheng et
al. 2012; Song et al. 2015). Panels (e) and (f) display the
observations of COR2, which also show the LF and the ejecta/HC
obviously. (See the animation (b) accompanying Figure 2 for the
eruption process). Based on the images recorded by EUVI, COR1 and
COR2, we conclude that the ejecta of this CME is a high
temperature and high density structure, which supports that the
ejecta observed by \textit{STEREO} is the HC recorded by
\textit{SDO}.

Associated with the CME, an obvious shock is generated as depicted
with the arrows in Figure 2(f). Generally, the diffuse front ahead
of the LF is interpreted as a shock structure (e.g., Vourlidas et
al. 2003, 2013; Feng et al. 2012, 2013), and the region between
the shock and the HC is usually termed as the sheath. The shock,
sheath, and ejecta observed in COR2 images was detected with the
in-situ observations sequentially as displayed in next subsection.

\subsection{The in-situ detection of the ICME}
Figure 3 shows the in-situ measurements from the SWEPAM, MAG, and
SWICS on board \textit{ACE} at the Lagrangian point (L1). The
normalized pitch angle distribution (PAD) of 272 eV electrons
(panel a), the solar wind speed (black line) and ratios of its
three components to the total speed (panel b), magnetic field
strength (black line) and its three components (panel c), proton
density and temperature (panel d), plasma $\beta$ and total
pressure (panel e), and the entropy and average Fe charge state
(panel f) are displayed from top to bottom. Note the velocity
(panel b) and magnetic field (panel c) components are plotted in
the GSE (Geocentric Solar Ecliptic) coordinate, where X-axis (red
line) points from the Earth towards the Sun, Y-axis (green line)
is chosen to be in the ecliptic plane pointing towards dusk
(opposite the Earth's motion), and Z-axis (blue line) is parallel
to the ecliptic pole.

The shock, sheath, and ejecta (the shaded region) appeared
sequentially as expected above. The ejecta part is an MC, with
rotations of magnetic field components (especially By, the green
line), low temperature and density, as well as low plasma $\beta$
compared to the background solar wind. Therefore, the ejecta
should be an MFR structure, which is also supported by the
bidirectional electrons (BDEs) as shown in panel (a), especially
in the former part of the MC, because both footpoints of the MFR
still anchor on the Sun and the obvious signature of BDEs will
appear when a spacecraft passes (e.g., Kilpua et al. 2013; Song et
al. 2015).

The average Fe charge state in the MC is apparently elevated
compared to the background solar wind and the sheath region as
shown in panel (f) with the red line, which is not an uncommon
phenomenon for MCs (e.g., Kilpua et al. 2013). The iron charge
distribution was even used as an identifier of ICMEs (Lepri et al.
2001), because the high ionized iron was produced in the current
sheet connecting the MFR/HC and the flare loops through magnetic
reconnection, and then filled in the MFR/HC (e.g., Ko et al.
2013). Therefore, the ion charge state offers an important clue to
relate the MC and HC as its distribution is fully established
within a few solar radii from the Sun and remains frozen after
that (e.g., Esser \& Edgar 2001; Chen et al. 2004).

As described above, the erupted HC existed prior to the flare
onset with a relative lower temperature ($\sim$5 MK). As the
reconnection takes place along the post-CME current sheet, the
pre-existing HC structure will be added more layers of plasma with
the reconnected field lines, and the heated plasma will fill in
the HC structure like `layers of an onion' (Lin et al. 2004; Ko et
al. 2013). Therefore, we would expect to observe an MC with a
low-ionization-state center and a high-ionization-state shell for
this event, which is confirmed by the observations as shown in
panel (f). The average Fe charge state near the MC center is close
to 10.5+, apparently lower than those at the shell (beyond 12+),
but higher slightly than the background solar wind (below 10+) as
the HC prior to the eruption had a higher temperature ($\sim$5 MK)
than the background solar wind ($\sim$1-2 MK) in the corona. Note
the background average Fe charge state is depicted with the blue
dotted line. For the filaments contained in ICMEs, their in-situ
average ion charge states might be lower than those in the
background solar wind (Lepri \& Zurbuchen 2010; Ko et al. 2013) as
their initial temperatures in the corona are lower than their
backgrounds, which does not conflict with our observations. Our
average Fe charge state observations are well consistent with the
expectation qualitatively, which provides a further support that
the MC is the interplanetary counterpart of the erupted HC in the
corona for this event. Note the pre-existing HC will grow up and
keep its high temperature during the eruption, and the MC
corresponds to the final HC after the flare reconnection.
Therefore, the pre-existing HC should correspond to the central
part of the MC. Unfortunately, as the SWICS on board \textit{ACE}
suffered a hardware anomaly in 2011 and lost the ability to
provide reliable iron charge state distributions, no quantitative
temperature analysis is provided based on the charge state
distributions.

\section{DISCUSSION}
Song et al. (2015) identified the counterpart of an HC in
interplanetary space for the first time through analyzing its high
temperature, appearance behind the shock and sheath, and the
associated BDEs. They suggested that the HC will not evolve into a
typical MC with low temperature under some special conditions. For
instance, if there exists a corotating interaction region (CIR)
ahead of the HC, then the CIR can form a magnetic container to
inhibit the expansion of the HC and cool it down to a low
temperature, which is the case reported in Song et al. (2015).
They also showed that the spacecraft passed far away from the HC
center, so no regular rotations of magnetic field components were
observed. Therefore, they did not give the strong evidence to
support that the HC is the MFR. However, the present event shows a
pure ICME event, and there is not a CIR structure interacting with
the ICME according to the in-situ observations. Therefore, the HC
can expand freely and evolve into a low temperature and low
density structure during its propagation to $\sim$1 AU.

MCs are detected in only about 30\% of ICMEs (Richardson \& Cane
2010; Wu \& Lepping 2011). Riley \& Richardson (2013) summarized
the factors to explain why some ICMEs are observed to be MCs and
others are not, including (1) the observational selection effect
of ICMEs, (2) the different initiation mechanisms of CMEs, (3) the
interactions of an MFR with itself or between neighboring MFRs,
and (4) the different evolutionary processes of MFRs. Based on our
present study and Song et al. (2015), we support that the
evolutionary process plays an important role on whether an HC will
form an MC during its propagation to $\sim$1 AU, and the process
can be greatly influenced by the interactions between the CME and
CIR.

Song et al. (2014b) demonstrated that an HC's temperature was
around 5 MK prior to the eruption and increased to $\sim$9 MK at
the flare peak, accompanying the growth of its volume. We suggest
that its heating process is likely to wear a higher temperature
``coat" for the pre-existing HC through magnetic reconnection. The
pre-existing HC might keep its original relative lower
temperature. Ciaravella \& Raymond (2008) pointed out that the
temperature in the current sheet can reach a maximum value of
$\sim$8 MK in an event, approaching to the coat temperature of the
HC and supporting our explanation. The low-temperature ``body" and
high-temperature ``coat" of the erupted HC might be corresponding
to the low-ionization-state center and high-ionization-state shell
of the MC.

\section{SUMMARY}
An HC erupted on 2012 July 12, accompanying an X1.4 class SXR
flare and a CME. The high-temperature images of AIA showed that
the HC had existed prior to the eruption and recorded its whole
eruption process. The EUVI and COR on board \textit{STEREO}
recorded the eruption edge on. The low-temperature images of EUVI
only showed the cooler LF ahead of the ejecta, while COR images
can present both the LF and the ejecta. These are consistent with
the high temperature and high density of the HC structure and
support that the ejecta corresponds to the erupted HC. In the
meanwhile, the associated CME shock ahead of the sheath region and
the ejecta was identified through the COR2 images, and these three
structures were clearly detected with the in-situ data
sequentially when the associated ICME passed through \textit{ACE}.
The ejecta evolved into an MC, containing a low-ionization-state
center and a high-ionization-state shell, which was consistent
with the pre-existing MFR observation and its growth process
through magnetic reconnection. All of these observations support
that the MC structure detected by \textit{ACE} is the counterpart
of the erupted HC in the lower corona, and the pre-existing HC
corresponds to the central part of the MC structure. Therefore,
our study provided strong observational evidence that the HC in
the lower corona is an MFR.

\acknowledgments We are grateful to the referee, Lan Jian, Jason
Gilbert, Haimin Wang, Bo Li, Chenglong Shen, Lulu Zhao, and Liang
Zhao for their valuable comments and discussion. We acknowledge
the use of data from the \textit{SDO}, \textit{STEREO} and
\textit{ACE} missions. This work is supported by the 973 program
2012CB825601, NNSFC grants 41274177, 41274175, and 41331068. J.Z.
is supported by US NSF AGS-1249270 and NSF AGS-1156120.

\clearpage

\begin{figure}
\epsscale{1.0} \plotone{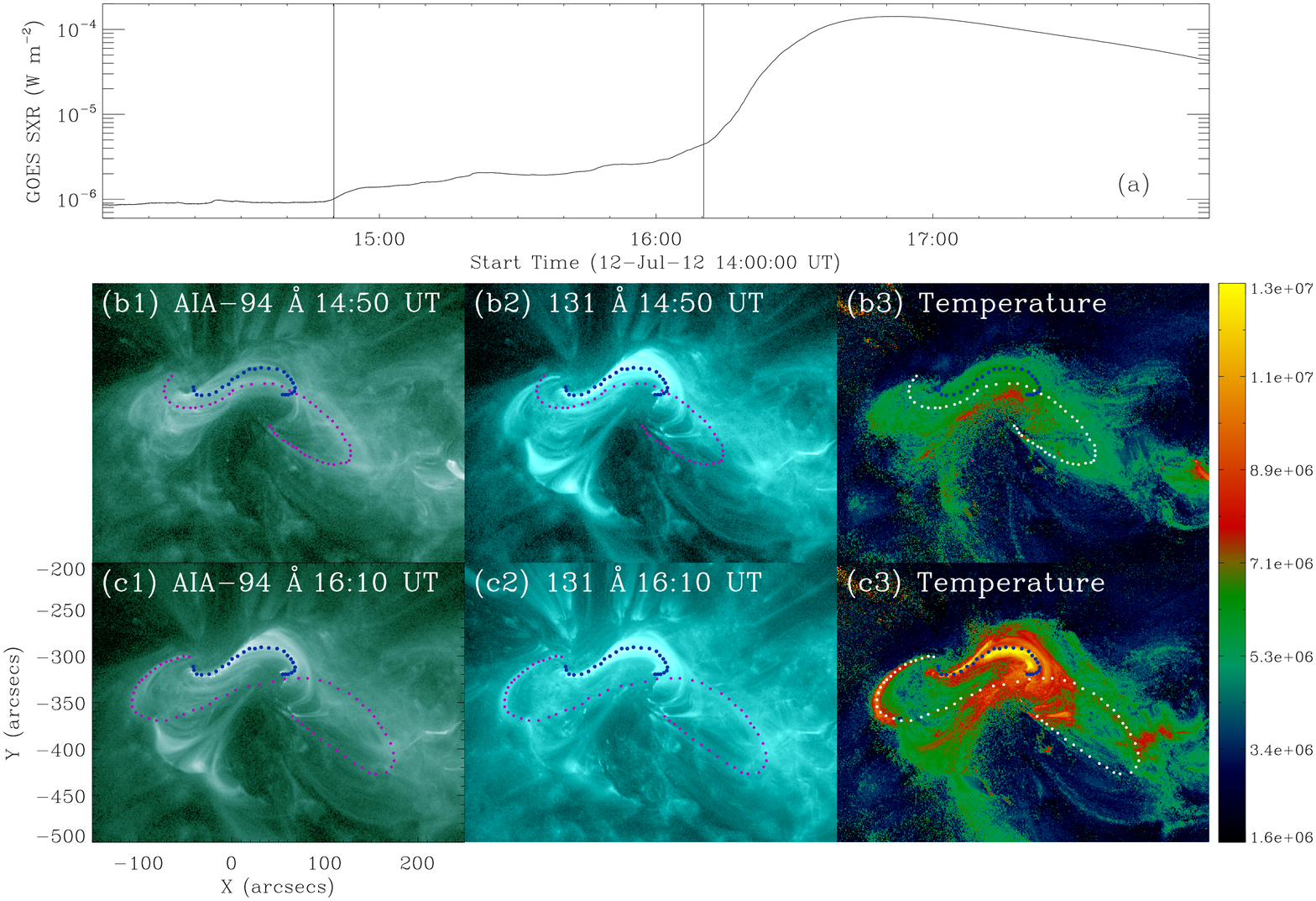} \caption{An HC eruption event on
2012 July 12. (a) The \textit{GOES} SXR 1-8 \AA \ flux profile of
the accompanying flare. (b1)-(b3) The AIA 94 \AA, 131 \AA, and
temperature image prior to the flare onset. (c1)-(c3) The same
with (b1)-(b3) but for a different time. (An animation of this
figure is available.) \label{Figure 1}}
\end{figure}

\begin{figure}
\epsscale{0.85} \plotone{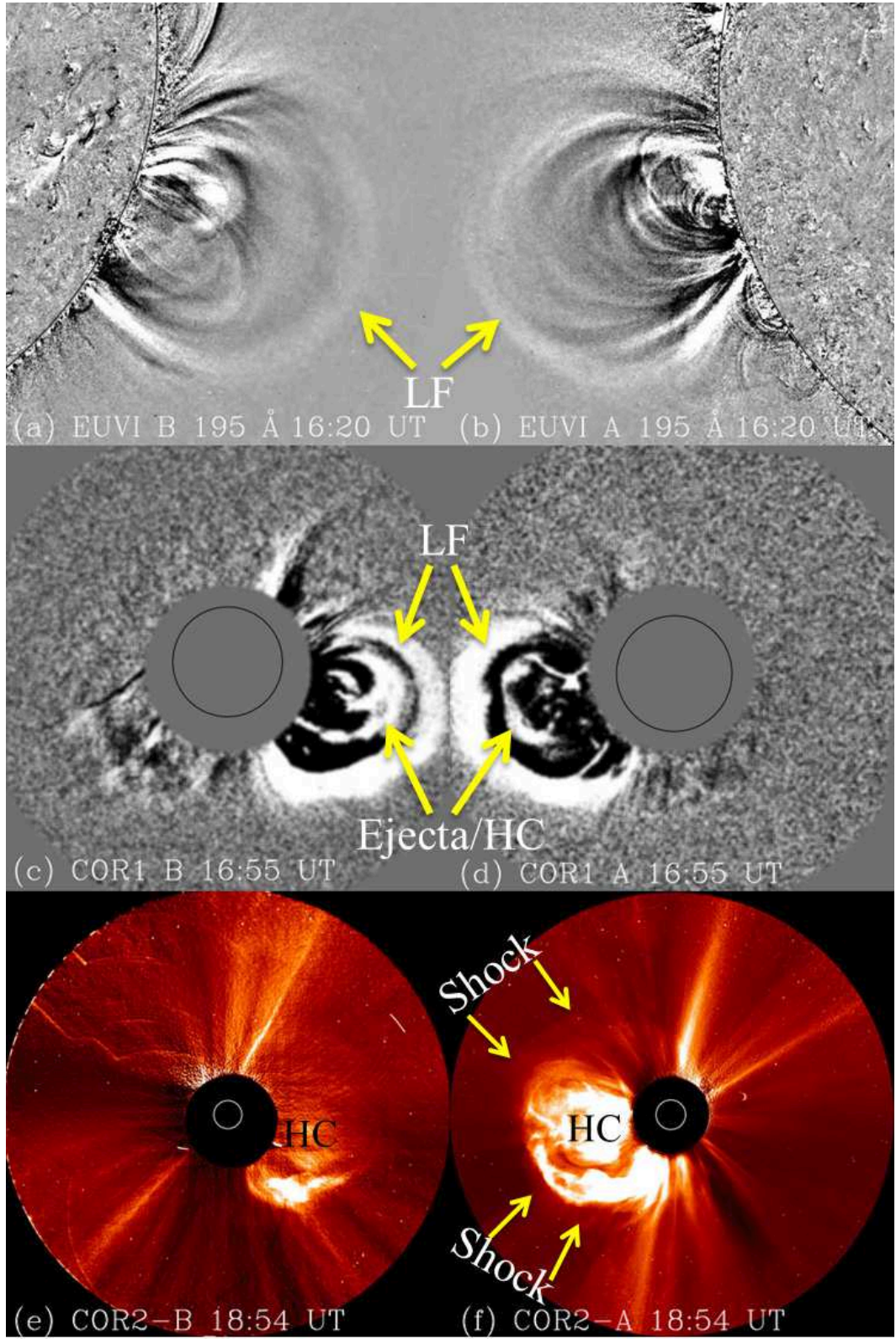} \caption{(a),(b) The EUVI
difference images of \textit{STEREO} B and A. (c),(d) The COR1
difference images of the CME. (e),(f) The COR2 direct images of
the CME. (Animations (a and b) of this figure are available.)
\label{Figure 2}}
\end{figure}

\begin{figure}
\epsscale{0.85} \plotone{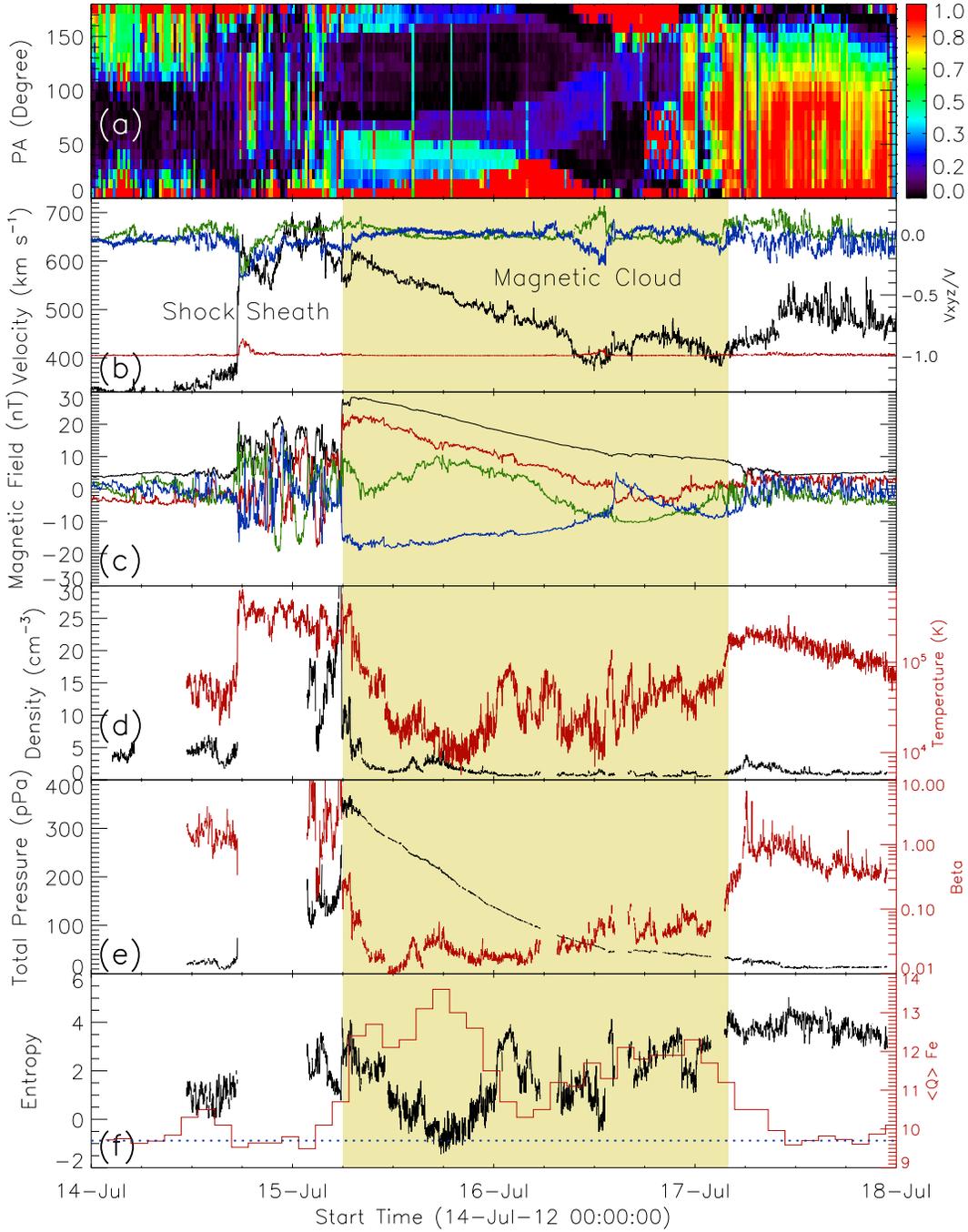} \caption{Solar wind parameters
measured with \textit{ACE}. From top to bottom, the panels show
the PAD of electrons at 272 eV, bulk speed, magnetic field,
density and temperature, plasma $\beta$ and total pressure, as
well as entropy and average Fe charge state. \label{Figure 3}}
\end{figure}


\begin{thebibliography}{}

\bibitem[Burlaga et al.(1981)]{1981JGR....86.6673B} Burlaga, L., Sittler,
E., Mariani, F., \& Schwenn, R.\ 1981, \jgr, 86, 6673


\bibitem[Chen(2011)]{2011LRSP....8....1C} Chen, P.~F.\ 2011, Living Reviews
in Solar Physics, 8, 1


\bibitem[Chen et al.(2004)]{2004ApJ...602..415C} Chen, Y., Esser, R.,
Strachan, L., \& Hu, Y.\ 2004, \apj, 602, 415


\bibitem[Cheng et al.(2014)]{2014ApJ...780...28C} Cheng, X., Ding, M.~D.,
Guo, Y., et al.\ 2014a, \apj, 780, 28


\bibitem[Cheng et al.(2014)]{2014ApJ...789L..35C} Cheng, X., Ding, M.~D.,
Zhang, J., et al.\ 2014b, \apjl, 789, L35


\bibitem[Cheng et al.(2014)]{2014ApJ...789...93C} Cheng, X., Ding, M.~D.,
Zhang, J., et al.\ 2014c, \apj, 789, 93

\bibitem[Cheng et al. (2013)]{Cheng 2013}
Cheng, X., Zhang, J., Ding, M.D., Liu, Y., \& Poomvises, W.\ 2013,
\apj, 763, 43

\bibitem[Cheng et al.(2011)]{2011ApJ...732L..25C} Cheng, X., Zhang, J.,
Liu, Y., \& Ding, M.~D.\ 2011, \apjl, 732, L25


\bibitem[Cheng et al.(2012)]{2012ApJ...761...62C} Cheng, X., Zhang, J.,
Saar, S.~H., \& Ding, M.~D.\ 2012, \apj, 761, 62


\bibitem[Ciaravella
\& Raymond(2008)]{2008ApJ...686.1372C} Ciaravella, A., \& Raymond,
J.~C.\ 2008, \apj, 686, 1372


\bibitem[Del Zanna et
al.(2011)]{2011A&A...535A..46D} Del Zanna, G., O'Dwyer, B., \&
Mason, H.~E.\ 2011, \aap, 535, A46


\bibitem[Esser
\& Edgar(2001)]{2001ApJ...563.1055E} Esser, R., \& Edgar, R.~J.\
2001, \apj, 563, 1055


\bibitem[Feng et al.(2013)]{2013ApJ...767...29F} Feng, S.~W., Chen, Y.,
Kong, X.~L., et al.\ 2013, \apj, 767, 29


\bibitem[Feng et al.(2012)]{2012ApJ...753...21F} Feng, S.~W., Chen, Y.,
Kong, X.~L., et al.\ 2012, \apj, 753, 21


\bibitem[Gloeckler et al.(1998)]{1998SSRv...86..497G} Gloeckler, G., Cain,
J., Ipavich, F.~M., et al.\ 1998, \ssr, 86, 497


\bibitem[Gosling et al.(1991)]{1991JGR....96.7831G} Gosling, J.~T.,
McComas, D.~J., Phillips, J.~L., \& Bame, S.~J.\ 1991, \jgr, 96,
7831


\bibitem[Hess \& Zhang(2014)]{2014ApJ...792...49H} Hess, P., \& Zhang, J.\ 2014,
\apj, 792, 49


\bibitem[Howard et al.(2008)]{2008SSRv..136...67H} Howard, R.~A., Moses,
J.~D., Vourlidas, A., et al.\ 2008, \ssr, 136, 67

\bibitem[Kilpua et al. (2013)]{Kilpua 2013}
Kilpua, E. K. J., Isavnin, A., Vourlidas, A., Koskinen, H. E. J.,
\& Rodriguez, L. 2013, Ann. Geophys., 31, 1251

\bibitem[Ko et al.(2013)]{2013AIPC.1539..207K} Ko, Y.-K., Raymond, J.~C.,
Rakowski, C., \& Rouillard, A.\ 2013, American Institute of
Physics Conference Series, 1539, 207


\bibitem[Lemen et al.(2012)]{2012SoPh..275...17L} Lemen, J.~R., Title,
A.~M., Akin, D.~J., et al.\ 2012, \solphys, 275, 17


\bibitem[Lepri et al.(2001)]{2001JGR...10629231L} Lepri, S.~T., Zurbuchen,
T.~H., Fisk, L.~A., et al.\ 2001, \jgr, 106, 29231

\bibitem[Lepri \& Zurbuchen (2010)]{Lepri 2010}
Lepri, S. T., \& Zubruchen, T. H. 2010, \apj, 723, L22

\bibitem[Lin et al.(2004)]{2004ApJ...602..422L} Lin, J., Raymond, J.~C.,
\& van Ballegooijen, A.~A.\ 2004, \apj, 602, 422


\bibitem[M{\"o}stl et al.(2014)]{2014ApJ...787..119M} M{\"o}stl, C., Amla,
K., Hall, J.~R., et al.\ 2014, \apj, 787, 119


\bibitem[McComas et al.(1998)]{1998SSRv...86..563M} McComas, D.~J., Bame,
S.~J., Barker, P., et al.\ 1998, \ssr, 86, 563


\bibitem[McKenzie
\& Canfield(2008)]{2008A&A...481L..65M} McKenzie, D.~E., \&
Canfield, R.~C.\ 2008, \aap, 481, L65


\bibitem[O'Dwyer et
al.(2010)]{2010A&A...521A..21O} O'Dwyer, B., Del Zanna, G., Mason,
H.~E., Weber, M.~A., \& Tripathi, D.\ 2010, \aap, 521, A21


\bibitem[Patsourakos et al.(2013)]{2013ApJ...764..125P} Patsourakos, S.,
Vourlidas, A., \& Stenborg, G.\ 2013, \apj, 764, 125


\bibitem[Richardson
\& Cane(2010)]{2010SoPh..264..189R} Richardson, I.~G., \& Cane,
H.~V.\ 2010, \solphys, 264, 189


\bibitem[Riley
\& Richardson(2013)]{2013SoPh..284..217R} Riley, P., \&
Richardson, I.~G.\ 2013, \solphys, 284, 217


\bibitem[Rust
\& Kumar(1994)]{1994SoPh..155...69R} Rust, D.~M., \& Kumar, A.\
1994, \solphys, 155, 69


\bibitem[Shen et al.(2014)]{2014JGRA..119.7128S} Shen, F., Shen, C., Zhang,
J., et al.\ 2014, Journal of Geophysical Research (Space Physics),
119, 7128


\bibitem[Smith et al.(1998)]{1998SSRv...86..613S} Smith, C.~W., L'Heureux,
J., Ness, N.~F., et al.\ 1998, \ssr, 86, 613


\bibitem[Song et al.(2014)]{2014ApJ...792L..40S} Song, H.~Q., Zhang, J.,
Chen, Y., \& Cheng, X.\ 2014a, \apjl, 792, L40


\bibitem[Song et al.(2015)]{2015ApJ...803...96S} Song, H.~Q., Zhang, J.,
Chen, Y., et al.\ 2015, \apj, 803, 96


\bibitem[Song et al.(2014)]{2014ApJ...784...48S} Song, H.~Q., Zhang, J.,
Cheng, X., et al.\ 2014b, \apj, 784, 48




\bibitem[Titov
\& D{\'e}moulin(1999)]{1999A&A...351..707T} Titov, V.~S., \&
D{\'e}moulin, P.\ 1999, \aap, 351, 707


\bibitem[Vourlidas et al.(2013)]{2013SoPh..284..179V} Vourlidas, A., Lynch,
B.~J., Howard, R.~A., \& Li, Y.\ 2013, \solphys, 284, 179


\bibitem[Vourlidas et al.(2003)]{2003ApJ...598.1392V} Vourlidas, A., Wu,
S.~T., Wang, A.~H., Subramanian, P., \& Howard, R.~A.\ 2003, \apj,
598, 1392


\bibitem[Wang
\& Stenborg(2010)]{2010ApJ...719L.181W} Wang, Y.-M., \& Stenborg,
G.\ 2010, \apjl, 719, L181


\bibitem[Wu
\& Lepping(2011)]{2011SoPh..269..141W} Wu, C.-C., \& Lepping,
R.~P.\ 2011, \solphys, 269, 141


\bibitem[Zhang et al.(2003)]{2003ApJ...582..520Z} Zhang, J., Dere, K.~P.,
Howard, R.~A., \& Bothmer, V.\ 2003, \apj, 582, 520


\bibitem[Zhang et al.(2007)]{2007JGRA..11210102Z} Zhang, J., Richardson,
I.~G., Webb, D.~F., et al.\ 2007, Journal of Geophysical Research
(Space Physics), 112, A10102


\bibitem[Zhang et al.(2012)]{2012NatCo...3E.747Z} Zhang, J., Cheng, X.,
\& Ding, M.-D.\ 2012, Nature Communications, 3, 747



\end{thebibliography}
\end{document}